\documentclass[conference]{IEEEtran}
\IEEEoverridecommandlockouts

\usepackage{cite}
\usepackage{booktabs}
\usepackage{amsmath,amssymb,amsfonts}
\usepackage{algorithmic}
\usepackage{graphicx}
\usepackage{textcomp}
\usepackage{xcolor}
\usepackage{pifont}
\newcommand{\xmark}{\ding{55}} 
\usepackage{makecell}
\usepackage{enumitem}
\usepackage{multirow}
\usepackage{tabularx}
\usepackage[most]{tcolorbox}
\usepackage{hyperref}
\usepackage{balance}
\usepackage{flushend}

\begin{document}

\IEEEoverridecommandlockouts
\IEEEpubid{\makebox[\columnwidth]{979-8-3315-9147-2/25/\$31.00~\copyright~2025 IEEE\hfill}
\hspace{\columnsep}\makebox[\columnwidth]{}}

\title{Contextual Code Retrieval for Commit Message Generation: A Preliminary Study}

\author{
    \IEEEauthorblockN{Bo Xiong, Linghao Zhang, Chong Wang$^{*}$\thanks{\indent This work is funded by the NSFC with Grant No. 62172311, 62032016, and the numerical calculations in this paper have been done on the supercomputing system in the Supercomputing Center of Wuhan University.}, Peng Liang$^{*}$}
    \IEEEauthorblockA{School of Computer Science, Wuhan University, Wuhan, China}
    \IEEEauthorblockA{\{\href{mailto:yueshaomoon_@whu.edu.cn}{yueshaomoon\_},\href{mailto:starryzhang@whu.edu.cn}{starryzhang}, \href{mailto:cwang@whu.edu.cn}{cwang}, \href{mailto:liangp@whu.edu.cn}{liangp}\}@whu.edu.cn}
}

\maketitle

\IEEEpubidadjcol

\begin{abstract}
A commit message describes the main code changes in a commit and plays a crucial role in software maintenance. Existing commit message generation (CMG) approaches typically frame it as a direct mapping which inputs a code diff and produces a brief descriptive sentence as output. However, we argue that relying solely on the code diff is insufficient, as raw code diff fails to capture the full context needed for generating high-quality and informative commit messages. In this paper, we propose a contextual code retrieval-based method called C3Gen to enhance CMG by retrieving commit-relevant code snippets from the repository and incorporating them into the model input to provide richer contextual information at the repository scope. In the experiments, we evaluated the effectiveness of C3Gen across various models using four objective and three subjective metrics. Meanwhile, we design and conduct a human evaluation to investigate how C3Gen-generated commit messages are perceived by human developers. The results show that by incorporating contextual code into the input, C3Gen enables models to effectively leverage additional information to generate more comprehensive and informative commit messages with greater practical value in real-world development scenarios. Further analysis underscores concerns about the reliability of similarity-based metrics and provides empirical insights for CMG.
\end{abstract}


\section{Introduction}
In software development and maintenance, Git has become the most widely used distributed version control system (VCS), where commits serve as the fundamental unit for recording the evolution of code. Each commit typically consists of two components: a code diff and a commit message. The commit message is intended to describe the nature, motivation, and potential impact of the corresponding code changes. Normally, a good commit message provides a concise, one-sentence summary of the change and clearly conveys the rationale behind the change. However, composing informative commit messages can be both time-consuming and labor-intensive for developers. Many find the process tedious and lack sufficient motivation to invest effort in writing them. As a result, in real-world development settings, the overall quality of commit messages is often suboptimal. A recent study~\cite{tian2022makes} reports that approximately 44\% of the commit messages fail to meet quality expectations, indicating that many messages lack essential information and fail to convey key details about the changes and their underlying rationale.

Against this backdrop, CMG has emerged as a prominent topic in automated software engineering, attracting growing attention from the research community. Existing approaches can be broadly categorized into three groups: retrieval-based methods, learning-based methods, and hybrid methods. With recent advances in large language models (LLMs), these models have been increasingly applied to both natural language processing and code-related tasks. Several studies~\cite{zhang2024automatic,zhang2024using} have begun to explore their potential in CMG tasks and have reported promising prospects.

Most existing CMG approaches formulate the task as a direct mapping $f$ that maps a code diff to a message, i.e., $f(\text{diff}) \rightarrow \text{message}$. These approaches mainly focus on optimizing the architecture of the model $f$ itself. However, for CMG tasks, it is insufficient to rely solely on the information extracted from the code diff, since the \texttt{.diff} file in Git only contains a limited window of code around the changes, lacking access to a broader knowledge of the cross-files and repository. Instead, attention should also be directed toward augmenting the input context, so that the model can generate commit messages with richer semantics to better capture the global development intent. For this purpose, this paper proposed C3Gen as a retrieval-augmented framework to enhance commit message generation by incorporating relevant contextual code snippets alongside the code diff, in order to improve the quality of the automatically generated commit messages.


The main contributions of this paper include: (1) the construction of a high-quality and large-scale dataset of commits, ApacheCM, which comprises over 230,000 commits from Apache open-source projects and serves as a valuable resource for CMG research; (2) the design of C3Gen, a retrieval-augmented framework that enhances CMG by retrieving and incorporating commit-relevant code snippets from the repository to enrich the model input; and (3) extensive experiments conducted across various large language models, where we evaluate the effectiveness of C3Gen using both objective metrics and an empirical human evaluation, offering a comprehensive understanding of the practical quality and usability of the C3Gen-generated commit messages.

The replication package of this paper is available at~\cite{replication}, including the code and data.
\begin{figure*}[!t]
    \centering
    \includegraphics[width=0.8\linewidth]{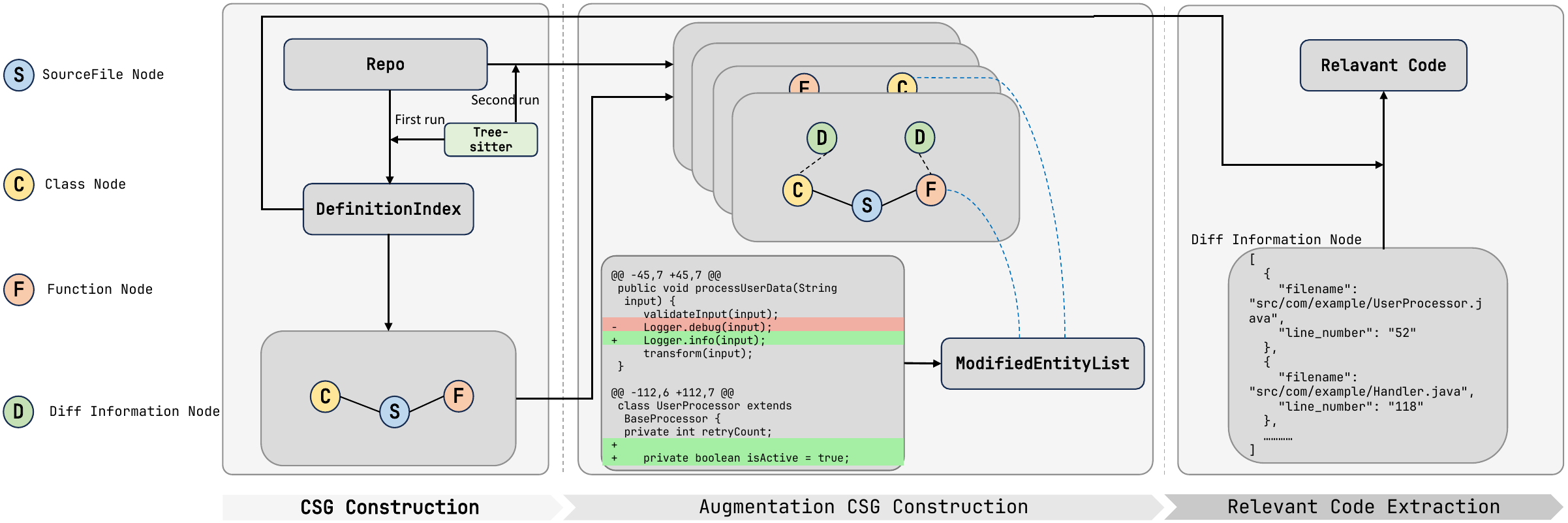}
    \caption{Overview of the C3Gen Framework}
    \label{fig: overview}
\end{figure*}

\section{Related Work}


Researchers have proposed various approaches for CMG, most of which either solely rely on raw code diffs as input or employ different generation mechanisms to produce commit messages. These methods can be broadly categorized into the following three groups:
(1) \textbf{Rule-based approaches}~\cite{buse2010automatically,linares2015changescribe,shen2016automatic} were used in CMG, where predefined rules or templates were utilized to construct sentences that describe the changes in the code diff. (2) \textbf{Retrieval-based approaches} leverage information
retrieval techniques to suggest commit messages from similar code diffs. NNGen~\cite{liu2018neural} computes cosine similarity between the target code diff and those in the corpus, selects the top-k diff-message pairs, and uses the commit message of the one with the highest BLEU score as the output. (3) \textbf{Learning-based approaches} typically treat CMG as a Neural Machine Translation (NMT) task. These methods leverage large-scale diff-message datasets to train deep neural networks capable of learning how to generate commit messages. CommitGen~\cite{jiang2017automatically} represents an early attempt to apply NMT to CMG. It trained a recurrent neural network with an encoder-decoder architecture, yielding promising results.

In recent years, LLMs have demonstrated remarkable capabilities in text understanding and generation. Several studies have explored the use of LLMs for CMG~\cite{zhang2024using} , revealing impressive performance. Recently, several studies have explored the application of retrieval-augmented generation (RAG) to CMG, aiming to enrich the model input beyond the raw code diff. For instance, REACT~\cite{zhang2024rag} enhances the input context by retrieving similar historical commits and leverages LLMs’ in-context learning capabilities to improve generation performance. However, none of these approaches incorporates commit-relevant code snippets to provide the model with repository-level code context, which we argue is important for capturing the global scope of changes and the underlying development intent.

\section{Methodology}

To improve the quality of generated commit messages, we propose a \textbf{C}ontextual \textbf{C}ode-based approach for \textbf{C}ommit Message \textbf{Gen}eration, called \textbf{C3Gen} for short. 
C3Gen retrieves code snippets relevant to the commit by identifying program entities that invoke or instantiate the modified functions or classes, thereby incorporating richer contextual information into the generation process. As illustrated in Figure~\ref{fig: overview}, the C3Gen framework comprises three main stages: (1) constructing Code Structure Graphs (CSGs), (2) enriching the CSG with diff-related modifications, and (3) extracting contextually relevant code snippets for the CMG tasks later.

\label{sec:method}
\subsection{Stage I: CSG Construction}
In this stage, the \texttt{tree-sitter} parsing library is used to apply static analysis on each codebase to construct CSGs of sourcecode files included in the codebase. Unlike ASTs, which are used to capture detailed syntax of code, CSGs focus on high-level structures (classes and functions) to support efficient context retrieval. 

First, \texttt{tree-sitter} is invoked to parse the definitions of all classes and functions in each source code file, in order to extract the names of classes or functions, their file paths, as well as the start and end lines of the class or function definitions. These elements extracted from each sourcecode file are serialized and stored in one JSON file, and the JSON files of all sourcecode files in the codebase are referred to \texttt{DefinitionIndex}. 

Second, the CSG of each source code file in the codebases is constructed based on the corresponding JSON file in \texttt{DefinitionIndex}. More specifically, the root node (S-node) of a CSG refers to a specified source code file. The remaining nodes in this CSG, i.e., F-node and C-node,  denote the functions and classes in the corresponding JSON file, respectively. The edges between S-node and F-node/C-node denote that the functions and/or classes are defined in this source code file.  

\subsection{Stage II: Diff-based Augmentation of CSGs}

In Stage II, the initial CSGs of one codebase, which are created in Stage I, will be augmented with information related to code changes, which point to the specified code diff. 

For this purpose, the code diff is parsed and partitioned into several smaller segments, according to the different source code files to which these segments refer. Then, it is needed to identify the modified entities in each segment, including their names and types (i.e., function or class). After removing duplicates, these identified entities are represented as name-type pairs and stored in a list named \texttt{ModifiedEntityList}.

Then, the second-round parsing with \texttt{tree-sitter} is conducted on each source code file in the codebase to identify the source code files that are invoked by the functions or instantiate the classes defined in this source code file. If the names of these classes or functions match the ones in \texttt{ModifiedEntityList}, the corresponding C-nodes or F-nodes in the CSG of this source code file will be augmented with code diff information. The augmented information is stored in the linked D-nodes to cover the names and paths of the invoked source code files as well as the line number referring to the locations of instantiation, invocation, etc.

\subsection {Stage III: Relevant Code Snippet Extraction}

Stage III leverages the \texttt{ModifiedEntityList} and the augmented CSG to get relevant code context. For each entity in  \texttt{ModifiedEntityList}, we locate its corresponding information nodes in the CSG to retrieve all the recorded invocations or instantiations. If the invocation occurs within the definition of one function or class, the entire body of the enclosing function or class will be extracted by using the elements from \texttt{DefinitionIndex}. If the invocation is located in a global scope or unstructured block, 25 lines before and after the invocation will be selected to as a heuristic for contextual relevance of the specified code diff. 

The union of all extracted segments is treated as the \textbf{Relevant Code Context} corresponding to the commit message, and will be used to enrich the input to the commit message generation model.

By integrating repository-level code context, C3Gen enables the model to generate commit messages that are informed not only by the immediate code changes but also by their usage, dependencies, and interactions across the broader codebase. This contributes to more semantically accurate and context-aware natural language summaries of code commits.

\section{ApacheCM: A New Dataset of Commits}

 With the growing interest in automatic commit message generation, several publicly available datasets~\cite{liu2018neural, jiang2017automatically, xu2019commit, wang2021context} have been widely adopted in recent research. However, most of these datasets typically consist of code diffs and their corresponding commit messages, omitting critical metadata such as repository names, commit SHA values, timestamps, etc. This lack of contextual information restricts the applicability of these datasets to broader research scenarios. Furthermore, persistent issues with respect to data quality and construction methodologies continue to plague existing datasets. For example, Zhang et al. have revealed that specific datasets did not implement rigorous quality filtering on source repositories, resulting in commit messages that exhibit marked inconsistent clarity and adherence to conventions~\cite{zhang2024using}. In addition, these datasets often expose incomplete data fields, which undermines their overall quality and usability.
 
Given the limitations mentioned above, it is necessary and valuable to construct a high-quality dataset for the CMG tasks. For this purpose, we introduce \textbf{ApacheCM} (\textbf{Apache} Projects \textbf{C}ommit \textbf{M}essage Dataset), a new dataset for CMG tasks. ApacheCM is constructed not only to support the experiments conducted herein, but also to potentially serve a standardized and reproducible benchmark for future research in this field.


\subsection{Data Selection and Collection}
\textbf{ApacheCM} is constructed based on Apache-hosted repositories on GitHub 
Specifically, we select the top 50 open source repositories ranked by the number of stars within the organization, encompassing prominent projects such as \textit{Superset}, \textit{ECharts}, \textit{Spark}, et. These repositories span a wide range of mainstream programming languages, thus guaranteeing both the representativeness and the diversity of the dataset.
Furthermore, commits from the main branch of each repository since 2015 are crawled, and one individual commit is treated as the atomic data unit in ApacheCM. For each commit, the GitHub REST API is employed to extract a set of structured fields, including gitUrl, code diff, commit message, full name of the repository, commit SHA, author of the commit, file names that the commit changes, timestamp of the commit, Lines of Code (LoC), etc. 



\begin{table}[!t]
\centering
\caption{Filtering Criteria for ApacheCM Dataset}
\begin{tabularx}{0.48\textwidth}{@{}p{2.2 cm}X@{}}
\toprule
\textbf{Criterion} & \textbf{Description} \\
\midrule
Message Length & Keep commit messages between 5–50 words. \\
\midrule
Diff Size & Exclude commits with over 300 changed lines. \\
\midrule
File Type & Include commits modifying at least one file in common languages (e.g., \texttt{.py}, \texttt{.java}, \texttt{.js}, etc.). \\
\midrule
Bot Detection & Remove commits by bots (\texttt{[bot]} in author name). \\
\midrule
Reverts \& Merges & Filter out commits containing ``merge'' or ``revert''. \\
\bottomrule
\end{tabularx}
\label{tab:filtering_criteria}
\end{table}

\subsection{Data Quality Assurance}
To ensure the quality, usability, and adaptability to downstream tasks of the ApacheCM dataset, we define six filtering criteria to select high-quality commit messages. These criteria are summarized in Table~\ref{tab:filtering_criteria}.

After implementing the aforementioned six filtering criteria, \textbf{ApacheCM} dataset comprises 234,799 commits from 50 high-quality Apache open-source projects. We then randomly sample 10,000 commits to serve as the test set. Each commit in ApacheCM encompasses comprehensive metadata, code diff, and the corresponding commit messages. The comparison of ApacheCM and existing datasets is shown in Table~\ref{tab:datasets}.

\begin{table*}[!ht]
\centering
\caption{ApacheCM v.s. Existing Commit Message Datasets}
\label{tab:datasets}
\begin{tabular}{lccccc}
\toprule
\textbf{Dataset} & \textbf{CommitGen} & \textbf{NNGen} & \textbf{CoDiSum} & \textbf{MCMD} & \textbf{ApacheCM (Ours)} \\
\midrule
\#Train Set         & 26,208    & 22,112  & 75,000  & 1,800,000  & 249,830 \\
\#Validation Set    & 3,000     & 3,000   & 8,000   & 225,000    & 10,000 \\
\#Test Set          & 3,000     & 2,521   & 7,661   & 225,000    & 10,000 \\
\#Repositories      & 1,000     & 1,000   & 1,000   & 500        & 50\textsuperscript{*} \\
Filtering Steps     & \xmark    & \xmark  & \checkmark & \checkmark  & \checkmark \\
Deduplication       & \xmark    & \xmark  & \xmark      & \checkmark  & \checkmark \\
Reproducible        & \xmark    & \xmark  & \xmark      & \xmark      & \checkmark \\
Complete Fields     & \xmark    & \xmark  & \xmark      & \xmark      & \checkmark \\
Languages Covered   & Java      & Java    & Java        & \makecell[l]{Java, C\#, C++,\\ Python, JavaScript} & \makecell[l]{Java, C++, Python,\\and other 6 languages} \\
\bottomrule
\end{tabular}
\end{table*}
\section{Experimental Setup}

\subsection{Research Question}



This preliminary study intends to answer a single Research Question (\textbf{RQ}): \textbf{\textit{Compared to the baselines using only code diffs as input, does the integration of retrieved code snippets improve the performance of commit message generation models, in terms of both objective metrics (e.g., BLEU, ROUGE, METEOR) and human-judged subjective metrics (e.g., Clarity, Completeness, Correctness)?}} We want to know if the generator gets both the diff and its associated retrieved code snippets as input, will C3Gen perform better on the CMG task than when the model only has the code diff? This RQ aims to compare two input setups: one is code diff only, and another incorporates both the code diff and retrieved code snippets, across common objective and subjective evaluation metrics.

\subsection{Model Selection}
In our experiments, four representative LLMs, i.e., GPT-4o, GPT-4.1, DeepSeek V3, and DeepSeek R1, are selected to perform generation. These models cover both open-source (DeepSeek) and proprietary type (GPT), as well as general and reasoning model (DeepSeek R1). We access all models via their official APIs and uniformly set the temperature parameter to 0.0, which minimizes randomness.

\subsection{Metrics}
For a generated commit message, we quantify its generation quality using both objective similarity-based metrics and subjective metrics derived from human evaluation.

\subsubsection{Objective Metrics}
Following prior CMG work~\cite{dong2022fira,hoang2020cc2vec,xu2019commit}, four automatic evaluation metrics are employed to quantitatively measure the similarity between the commit messages automatically generated by LLMs and those written by human developers. Higher scores on these metrics denote a greater degree of similarity.

\paragraph{BLEU~\cite{papineni2002bleu}}
As a classical metric in machine translation, BLEU is used to evaluate textual similarity by analyzing n-gram co-occurrence statistics. In this study, we selected the \texttt{google\_bleu} implementation from HuggingFace.

\paragraph{ROUGE-L~\cite{lin2004rouge}}
This metric is used to evaluate sequence similarity based on the Longest Common Subsequence (LCS). Unlike n-gram matching, LCS does not require the matched words to be contiguous in the sequence but does require them to appear in the same order. This better captures the semantic order consistency.

\paragraph{METEOR~\cite{banerjee2005meteor}}
METEOR evaluates semantic similarity via flexible unigram alignment. It prioritizes recall to reward coverage and penalizes incoherent alignments to prevent score inflation from spurious matches.

\paragraph{CIDEr~\cite{vedantam2015cider}}
Originally developed for image captioning, CIDEr is also widely used for evaluating CMG. It quantifies similarity based on the consensus between generated text and human references, using TF-IDF weighting to highlight salient n-grams and better capture alignment with human.


\subsubsection{Subjective Metrics}
\label{sec:human}
To better understand human developers’ preferences toward generated commit messages, we provide three subjective metrics, i.e., \textit{Clarity}, \textit{Completeness}, and \textit{Correctness}. Each metric is scored on a 1–5 point scale, where a higher score indicates a better commit message. More specifically, \textit{Clarity} is defined to evaluate how easily the commit message can be understood, considering its wording, structure, and grammar. 
\textit{Completeness} assesses how thoroughly the commit message captures all changes in the code diff, including important and relevant contextual details.
\textit{Correctness} assesses how accurately the commit message reflects the actual code changes, ensuring it avoids hallucinations or misinterpretations.

\section{Results}
In our experiments, we design two configurations to evaluate the performance of C3Gen for CMG. In the first configuration, the raw code diff is directly fed into the LLM to generate commit messages. By following the framework in Figure~\ref{fig: overview}, the second configuration uses C3Gen to retrieve relevant code segments for the code diff as additional information to enrich the input context. This is based on the hypothesis that providing more contextual information will help the semantic-augmented generation of commit messages, allowing to better illustrate why the change was made and what was changed.

\subsection{Objective Metric Evaluation}
Table~\ref{tab:rq1_results} presents the comparative results of objective metric scores between the method using only raw code diffs as input (denoted as ``Naive'' method) and the one enhanced with our proposed C3Gen by adding code-related context. 

\begin{table}[h]
\caption{Objective Metric Scores}
\centering
\label{tab:rq1_results}
\begin{tabular}{llrrrrr}
\toprule
\multirow{2}{*}{\textbf{Model}} & \multirow{2}{*}{\textbf{Method}} & \multicolumn{4}{c}{\textbf{Metric Scores (\%)}} \\
\cmidrule(lr{3pt}){3-6}
& & \textbf{BLEU} & \textbf{Rouge-L} & \textbf{METEOR} & \textbf{CIDEr}\\
\midrule \midrule
\multirow{2}{*}{GPT-4o} & Naive & 9.12 & 21.08 & 18.51 & 7.12 \\
 & C3Gen & 9.34 & 21.63 & 18.97 & 7.22 \\
 \midrule
\multirow{2}{*}{GPT-4.1} & Naive & 9.38 & 20.86 & 18.93 & 6.84 \\
 & C3Gen & 8.84 & 20.97 & 19.60 & 6.55 \\
 \midrule
\multirow{2}{*}{DeepSeek V3} & Naive & 10.03 & 22.75 & 19.64 & 8.18 \\
 & C3Gen & 9.80 & 22.10 & 19.18 & 8.13 \\
 \midrule
\multirow{2}{*}{DeepSeek R1} & Naive & 9.87 & 22.96 & 20.56 & 7.81 \\
 & C3Gen & 10.04 & 22.52 & 20.41 & 8.00 \\
\bottomrule
\end{tabular}
\end{table}

We observe that incorporating additional relevant code information yields limited improvements in objective metric scores, with differences that are statistically insignificant across most models. For example, C3Gen improves the BLEU score for GPT-4o by approximately 2.4\% (from 9.12 to 9.34), while the CIDEr score increases slightly from 7.12 to 7.22. However, for GPT-4.1, while METEOR increases from 18.93 to 19.60, the BLEU score drops from 9.38 to 8.84, indicating a trade-off between metrics.
Similarly, DeepSeek R1 exhibits a modest improvement in BLEU (from 9.87 to 10.04) and CIDEr (from 7.81 to 8.00), but a small drop in Rouge-L and METEOR. In contrast, DeepSeek V3 experiences a slight decline in all metrics when augmented with C3Gen, including a 2.3\% decrease in the BLEU score.

These mixed results suggest that different metrics emphasize distinct aspects of similarity. It is important to note that all of these metrics approximate quality based on heuristic similarity to the ground truth( the developer-written commit message). While some benefit from the added context (e.g., semantic richness as reflected in METEOR), others such as BLEU, which are more sensitive to n-gram overlap, may not consistently capture the improvements brought by additional information. This highlights the limitations of relying solely on automatic metrics to assess semantic adequacy and human preference in the context of CMG tasks.

In some cases, C3Gen results in slight score degradation for certain models and metrics, for example, a slight decrease in BLEU score for DeepSeek V3. We identified several cases that shed light on the potential reasons behind this observation. One possible reason lies in the complexity of the code diff. A single diff may involve modifications across multiple functions or classes. In many cases, the original commit message primarily describes changes to a key function, whereas the retrieval module might retrieve relevant code based on other, less central modifications. This mismatch between the source of the retrieved code and the focus of the commit message can negatively impact the evaluation metrics. Conversely, when the retrieved code is highly relevant to the core modified function, the metric scores will be improved.

\begin{figure}[!t]
  \centering
  \includegraphics[width=0.96\linewidth]{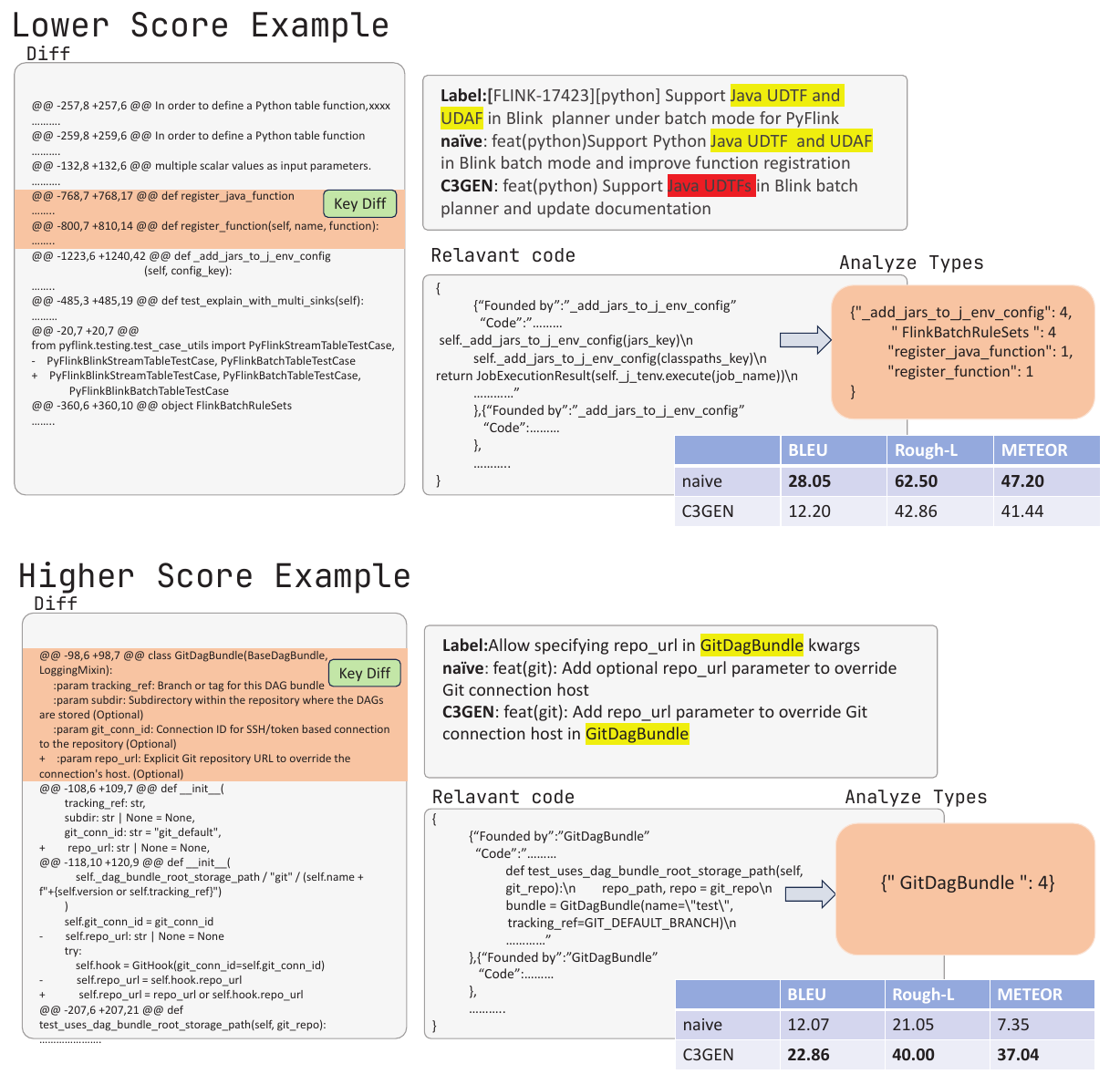}
  \caption{Examples with Higher and Lower Scores}
  \label{fig:Two Examples}
\end{figure}

More explanation of this reason can be provided by two representative examples shown in Figure~\ref{fig:Two Examples}. The lower score example demonstrates a scenario where the retrieved context is misaligned with the core modification in the diff, resulting in a suboptimal message. Among the ten retrieved code snippets, only two are directly related to the modified key function. In contrast, the higher score example shows a case where all retrieved contexts were closely tied to the core modification. A higher proportion of retrieved code snippets that are derived from the key modifications in the diff tends to improve the quality of the generated message significantly.

\begin{figure}[!b]
  \centering
  \includegraphics[width=0.9\linewidth]{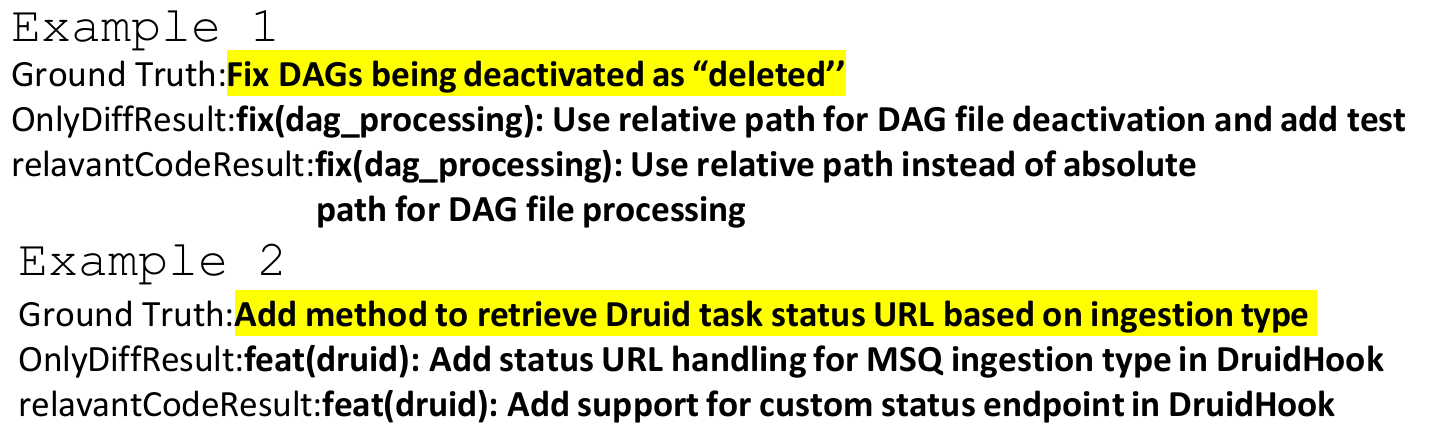}
  \caption{Two Examples of Stylistic Divergence}
  \label{fig:Stylistic Divergence}
\end{figure}

Another reason could be the stylistic divergence between ground truths and generated messages. As shown in Figure~\ref{fig:Stylistic Divergence}, each of the two instances includes three commits of one specific code diff: the ground truth is the one crawled from the repository, while OnlyDiffResult and relevantCodeResult refer to the generated commits with and without retrieved code-aware context respectively. It is observed that if the input is enriched with relevant codes, the generated messages may shift toward the variations with more (Example 1) or less (Example 2) implementation details. Therefore, a lower similarity between the ground truth and generated commits is caused by writing styles, rather than semantics. This indicates that current objective metrics may underestimate the quality of messages generated with semantic enhancement when their sylictic variations differ from the ground truth.


Furthermore, due to the inherent randomness in the output of LLMs, commit messages generated across different iterations may exhibit minor lexical variations while preserving semantic consistency. Generally, BLEU is designed to evaluate lexical overlap and is less sensitive to semantic equivalence. Therefore, a higher BLEU score will be given to the commit with more overlapping keywords, even if the semantics of generated messages are the same.  

Therefore, we argue that existing similarity-based objective metrics only provide heuristic approximations of generation quality. This calls for complementary human evaluation to gain deeper insights into the actual quality and semantic improvements of the generated commit messages.

\begin{tcolorbox}[
colback=gray!5,
colframe=black!50,
boxsep=0mm,
left=2mm, 
right=2mm
]
  \textbf{Key Finding 1:} C3Gen does not yield consistent improvements in similarity-based objective metric scores. 
 Further analysis reveals that these objective metrics only serve as heuristic approximations and cannot effectively capture the semantic enhancements brought about by including additional contextual code.
 \end{tcolorbox}

\subsection{Human Evaluation}
To more effectively assess the semantic quality of generated commit messages, beyond what automatic metrics can capture, we conduct a human evaluation on C3Gen-generated commit messages. We sampled a total of 370 instances from the test set based on a 95\% confidence level and a 0.05 margin of error. The evaluation was conducted by two authors of this paper, who served as participants and independently scored the generated commit messages. Both of them are majoring in Software Engineering and have over three years of programming experience. For each commit instance, participants were provided with the code diff and a set of nine commit message candidates generated by different methods, including the developer-written reference message. The order of the candidates was randomized to mitigate ordering bias. Each participant independently rated every commit message candidate across three subjective dimensions. The detailed scoring criteria, the instructions provided to the participants, and the resulting scores are included in~\cite{replication}.


\begin{table}[!b]
    \centering
    \caption{Results of Human Evaluation (Avg. of 2 participants)}
    \begin{tabular}{ccccc}
    \toprule
    \multicolumn{2}{c}{\textbf{Method}} & \textbf{Clarity} & \textbf{Completeness} & \textbf{Correctness}  \\
    \midrule
    \midrule
    \multicolumn{2}{c}{Reference} & 3.08 & 2.69 & 3.56 \\
    \midrule
    \multirow{2}{*}{GPT4o} & Naive & 4.36 & 3.98 & 4.63 \\
    & C3Gen & 4.06 & 4.16 & 4.61 \\
    \midrule
    \multirow{2}{*}{GPT4.1} & Naive & \textbf{4.46} & 4.43 & \textbf{4.87} \\
    & C3Gen & 4.22 & \textbf{4.50} & 4.77 \\
    \midrule
    \multirow{2}{*}{DeepSeek V3} & Naive & 3.94 & 3.51 & 4.30 \\
    & C3Gen & 3.63 & 3.75 & 4.32 \\
    \midrule
    \multirow{2}{*}{DeepSeek R1} & Naive & 3.84 & 3.57 & 4.30 \\
    & C3Gen & 3.56 & 3.76 & 4.26 \\
    \bottomrule
    \end{tabular}
    \label{tab:human}
\end{table}

Table~\ref{tab:human} presents the scores for the three subjective metrics, averaged across the two participants. We observed that for the \textit{Clarity} metric, incorporating additional code context led to a slight decrease in scores for the generated commit messages, suggesting that it may have increased the cognitive load on developers to some extent--likely due to the expanded scope and volume of information. The highest \textit{Clarity} score was achieved by GPT-4.1 when generating commit messages based solely on raw code diffs. In contrast, for the \textit{Completeness} metric, C3Gen consistently achieved substantial improvements across all models. This indicates that by incorporating relevant contextual code into the input, C3Gen enabled the models to effectively leverage the additional information, resulting in commit messages that conveyed more comprehensive and informative content. The \textit{Correctness} metric primarily assesses whether the commit message accurately aligns with the code diff without introducing misunderstandings. In this dimension, the performance of Naive and C3Gen is largely comparable.

In addition, it is interesting to observe that in human evaluation, the reference messages written by developers are treated as one of the candidates to be scored. They consistently received significantly lower scores across all evaluation metrics compared to the commit messages generated by LLMs. This suggests a conclusion that \textit{developer-written commit messages are of lower quality in terms of human preference}. This reinforces the concern that objective metrics based on similarity to reference messages may not accurately reflect the true quality of generated commit messages. More often, they serve only as heuristic indicators and are not reliable when used in isolation. An empirical human evaluation is therefore essential to understand how generated commit messages are perceived in real-world development scenarios.

\begin{tcolorbox}[
colback=gray!5,
colframe=black!50,
boxsep=0mm,
left=2mm, 
right=2mm
]
\textbf{Key Finding 2:} In human evaluation, the commit messages generated with C3Gen achieved around 5\% higher score on \textit{Completeness}. This indicates that additional code-related context can significantly enhance the informativeness of generated commit messages. However, there was a slight decrease in \textit{Clarity}, possibly due to increased complexity and cognitive load. The scores of \textit{Correctness} remain similar. In particular,cular, developer-written commit messages receive significantly lower scores across all dimensions, underscoring concerns about the reliability of similarity-based objective metrics.
\end{tcolorbox}

\section{Conclusions \& Future Work}

In this paper, we first construct a high-quality and context-rich dataset, i.e., ApacheCM, for CMG tasks. Unlike existing widely used datasets, ApacheCM includes comprehensive commit-level information, such as commit SHA, timestamp, and gitUrl, and is curated through a series of rigorous filtering strategies to ensure semantic clarity, relevance, and reproducibility. Then, we propose C3Gen as a retrieval-augmented framework to enhance CMG by effectively integrating relevant code snippets identified through advanced retrieval techniques. Our experimental results demonstrate that the proposed approach improves the performance of the CMG task. 

In the next step, we plan to further improve C3Gen by filtering the retrieved code snippets to retain only those that are directly related to the key diff, thereby enhancing the relevance of contextual information and reducing noise. 
Furthermore, we intend to extend ApacheCM by including more repositories and programming languages.

\bibliographystyle{IEEEtran}
\bibliography{cites}

\begin{thebibliography}{10}
\providecommand{\url}[1]{#1}
\csname url@samestyle\endcsname
\providecommand{\newblock}{\relax}
\providecommand{\bibinfo}[2]{#2}
\providecommand{\BIBentrySTDinterwordspacing}{\spaceskip=0pt\relax}
\providecommand{\BIBentryALTinterwordstretchfactor}{4}
\providecommand{\BIBentryALTinterwordspacing}{\spaceskip=\fontdimen2\font plus
\BIBentryALTinterwordstretchfactor\fontdimen3\font minus \fontdimen4\font\relax}
\providecommand{\BIBforeignlanguage}[2]{{%
\expandafter\ifx\csname l@#1\endcsname\relax
\typeout{** WARNING: IEEEtran.bst: No hyphenation pattern has been}%
\typeout{** loaded for the language `#1'. Using the pattern for}%
\typeout{** the default language instead.}%
\else
\language=\csname l@#1\endcsname
\fi
#2}}
\providecommand{\BIBdecl}{\relax}
\BIBdecl

\bibitem{tian2022makes}
Y.~Tian, Y.~Zhang, K.-J. Stol, L.~Jiang, and H.~Liu, ``What makes a good commit message?'' in \emph{Proceedings of the 44th International Conference on Software Engineering (ICSE)}.\hskip 1em plus 0.5em minus 0.4em\relax ACM, 2022, pp. 2389--2401.

\bibitem{zhang2024automatic}
Y.~Zhang, Z.~Qiu, K.-J. Stol, W.~Zhu, J.~Zhu, Y.~Tian, and H.~Liu, ``Automatic commit message generation: A critical review and directions for future work,'' \emph{IEEE Transactions on Software Engineering}, vol.~50, no.~4, pp. 816--835, 2024.

\bibitem{zhang2024using}
L.~Zhang, J.~Zhao, C.~Wang, and P.~Liang, ``Using large language models for commit message generation: A preliminary study,'' in \emph{Proceedings of the 31st IEEE International Conference on Software Analysis, Evolution and Reengineering (SANER)}.\hskip 1em plus 0.5em minus 0.4em\relax IEEE, 2024, pp. 126--130.

\bibitem{replication}
\BIBentryALTinterwordspacing
B.~Xiong, L.~Zhang, C.~Wang, and P.~Liang, ``Replication package of the paper: Contextual code retrieval for commit message generation: A preliminary study,'' May 2025. [Online]. Available: \url{https://doi.org/10.5281/zenodo.15502317}
\BIBentrySTDinterwordspacing

\bibitem{buse2010automatically}
R.~P. Buse and W.~R. Weimer, ``Automatically documenting program changes,'' in \emph{Proceedings of the 25th IEEE/ACM International Conference on Automated Software Engineering (ASE)}.\hskip 1em plus 0.5em minus 0.4em\relax ACM, 2010, pp. 33--42.

\bibitem{linares2015changescribe}
M.~Linares-V{\'a}squez, L.~F. Cort{\'e}s-Coy, J.~Aponte, and D.~Poshyvanyk, ``Changescribe: A tool for automatically generating commit messages,'' in \emph{Proceedings of the 37th IEEE/ACM International Conference on Software Engineering (ICSE)}.\hskip 1em plus 0.5em minus 0.4em\relax IEEE, 2015, pp. 709--712.

\bibitem{shen2016automatic}
J.~Shen, X.~Sun, B.~Li, H.~Yang, and J.~Hu, ``On automatic summarization of what and why information in source code changes,'' in \emph{Proceedings of the 40th IEEE Annual Computer Software and Applications Conference (COMPSAC)}.\hskip 1em plus 0.5em minus 0.4em\relax IEEE, 2016, pp. 103--112.

\bibitem{liu2018neural}
Z.~Liu, X.~Xia, A.~E. Hassan, D.~Lo, Z.~Xing, and X.~Wang, ``Neural machine-translation-based commit message generation: how far are we?'' in \emph{Proceedings of the 33rd ACM/IEEE International Conference on Automated Software Engineering (ASE)}.\hskip 1em plus 0.5em minus 0.4em\relax ACM, 2018, pp. 373--384.

\bibitem{jiang2017automatically}
S.~Jiang, A.~Armaly, and C.~McMillan, ``Automatically generating commit messages from diffs using neural machine translation,'' in \emph{Proceedings of the 32nd ACM/IEEE International Conference on Automated Software Engineering (ASE)}.\hskip 1em plus 0.5em minus 0.4em\relax IEEE, 2017, pp. 135--146.

\bibitem{zhang2024rag}
L.~Zhang, H.~Zhang, C.~Wang, and P.~Liang, ``Rag-enhanced commit message generation,'' \emph{arXiv preprint arXiv:2406.05514}, 2024.

\bibitem{xu2019commit}
S.~Xu, Y.~Yao, F.~Xu, T.~Gu, H.~Tong, and J.~Lu, ``Commit message generation for source code changes,'' in \emph{Proceedings of the 28th International Joint Conference on Artificial Intelligence (IJCAI)}.\hskip 1em plus 0.5em minus 0.4em\relax IJCAI, 2019, pp. 3975--3981.

\bibitem{wang2021context}
H.~Wang, X.~Xia, D.~Lo, Q.~He, X.~Wang, and J.~Grundy, ``Context-aware retrieval-based deep commit message generation,'' \emph{ACM Transactions on Software Engineering and Methodology}, vol.~30, no.~4, pp. 1--30, 2021.

\bibitem{dong2022fira}
J.~Dong, Y.~Lou, Q.~Zhu, Z.~Sun, Z.~Li, W.~Zhang, and D.~Hao, ``Fira: fine-grained graph-based code change representation for automated commit message generation,'' in \emph{Proceedings of the 44th International Conference on Software Engineering (ICSE)}.\hskip 1em plus 0.5em minus 0.4em\relax IEEE, 2022, pp. 970--981.

\bibitem{hoang2020cc2vec}
T.~Hoang, H.~J. Kang, D.~Lo, and J.~Lawall, ``Cc2vec: Distributed representations of code changes,'' in \emph{Proceedings of the 42nd ACM/IEEE International Conference on Software Engineering (ICSE)}.\hskip 1em plus 0.5em minus 0.4em\relax ACM, 2020, pp. 518--529.

\bibitem{papineni2002bleu}
K.~Papineni, S.~Roukos, T.~Ward, and W.-J. Zhu, ``Bleu: a method for automatic evaluation of machine translation,'' in \emph{Proceedings of the 40th Annual Meeting of the Association for Computational Linguistics (ACL)}.\hskip 1em plus 0.5em minus 0.4em\relax ACM, 2002, pp. 311--318.

\bibitem{lin2004rouge}
C.-Y. Lin, ``Rouge: A package for automatic evaluation of summaries,'' in \emph{Text Summarization Branches Out}, 2004, pp. 74--81.

\bibitem{banerjee2005meteor}
S.~Banerjee and A.~Lavie, ``Meteor: An automatic metric for mt evaluation with improved correlation with human judgments,'' in \emph{Proceedings of the ACL Workshop on Intrinsic and Extrinsic Evaluation Measures for Machine Translation and/or Summarization}.\hskip 1em plus 0.5em minus 0.4em\relax ACM, 2005, pp. 65--72.

\bibitem{vedantam2015cider}
R.~Vedantam, C.~Lawrence~Zitnick, and D.~Parikh, ``Cider: Consensus-based image description evaluation,'' in \emph{Proceedings of the IEEE Conference on Computer Vision and Pattern Recognition (CVPR)}.\hskip 1em plus 0.5em minus 0.4em\relax IEEE, 2015, pp. 4566--4575.

\end{thebibliography}

\balance

\end{document}